\documentclass{article}
\usepackage[T1]{fontenc}
\usepackage[utf8]{inputenc}
\usepackage{ismir}
\usepackage{amsmath,cite,url}
\usepackage{graphicx}
\usepackage{color}
\usepackage{amsfonts}
\usepackage{enumitem}
\usepackage{algorithm}
\usepackage{algorithmic}
\usepackage{amssymb}
\usepackage{booktabs}
\usepackage{pifont}

\usepackage{multirow}
\title{Rethinking Automatic Music Mixing as Sequential Stem Blending}

\multauthor
  {Yen-Tung Yeh$^1$ \hspace{1cm} Chung-Jui Chan$^1$ \hspace{1cm} Yun-Ning (Amy) Hung$^2$ \hspace{1cm} Yi-Hsuan Yang$^{1,3}$}
  {
  $^1$ GICE, National Taiwan University, \; 
  $^2$ Moises, \; 
  $^3$ AI-CoRE, National Taiwan University \; \\
  {\tt\small f12942179@ntu.edu.tw}
  }

\def\authorname{Yen-Tung Yeh, Chung-Jui Chan, Yun-Ning Hung, and Yi-Hsuan Yang}

\begin{document}

\maketitle

\begin{abstract}
Automatic music mixing, the task of automatically combining individual audio tracks into a cohesive mixture, is typically addressed by parallelized architectures that process all input tracks in a single pass. In this work, inspired by how human mix engineers process stems one at a time, we propose a paradigm shift and ask whether automatic music mixing can be reformulated as a sequential stem blending task, where each stem is blended into a growing submix. Specifically, we train a latent flow matching model conditioned on the submix context, enabling sequential processing of an arbitrary number of input tracks. To train the model, we introduce a degradation-based data synthesis strategy that simulates realistic stem blending scenarios from existing multitrack and source separation datasets. Experimental results on both stem blending and automatic music mixing benchmarks demonstrate the effectiveness of the proposed approach. We provide audio examples on the accompanying demo page\footnote{\url{https://sequential-mixing-demo.vercel.app/}}.
\end{abstract}

\section{Introduction}\label{sec:introduction}

\begin{figure}[!th]
 \centerline{
 \includegraphics[width=1.0\columnwidth, keepaspectratio]{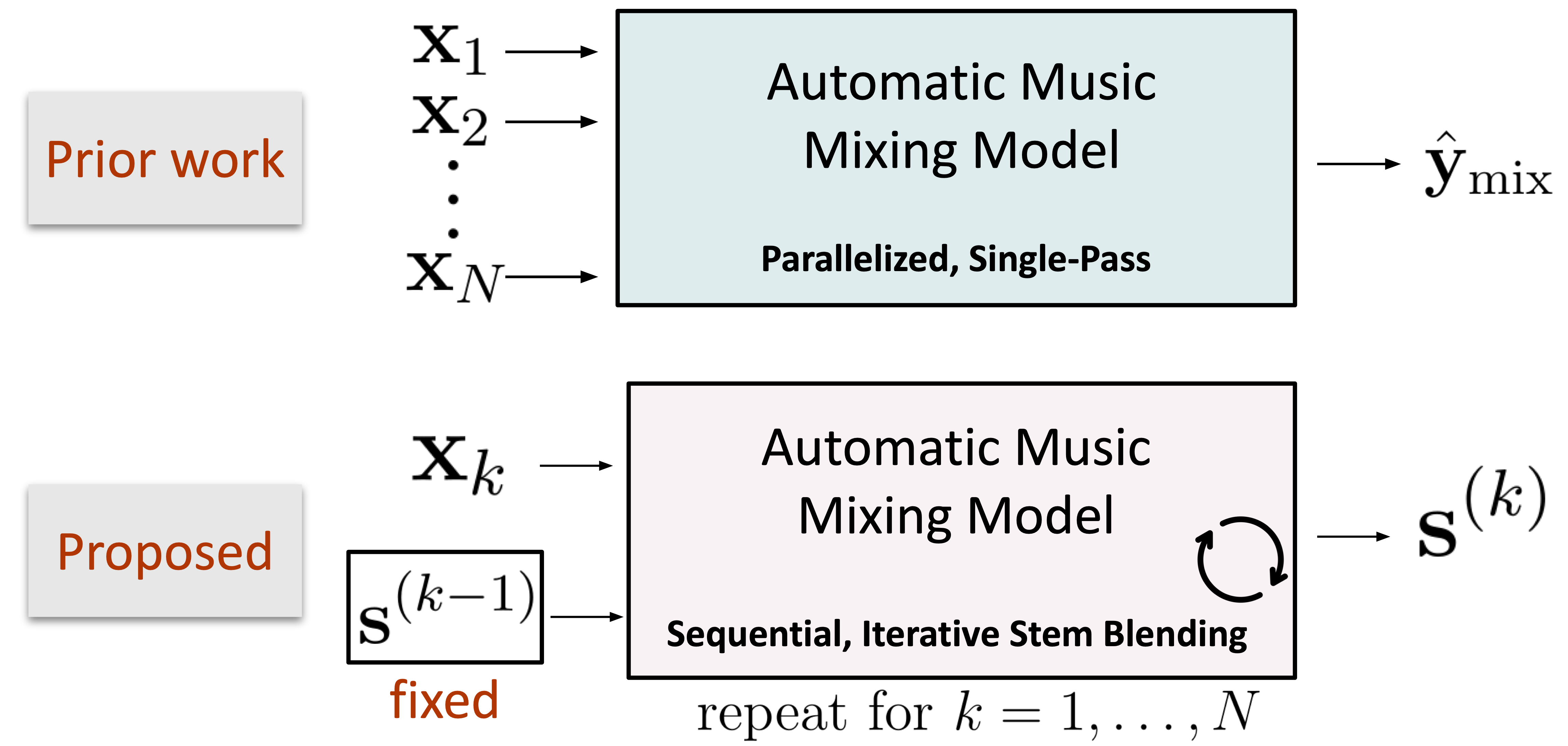}
 }
 \vspace{-2mm}
 \caption{Comparison of parallelized and sequential mixing paradigms. 
\textit{Prior work} (top) processes all $N$ stems simultaneously in a 
single pass to produce the final mixture $\hat{\mathbf{y}}_\text{mix}$. 
\textit{Proposed} (bottom) reformulates mixing as sequential stem 
blending, where a single stem $\mathbf{x}_k$ is integrated into a 
fixed submix $\mathbf{s}^{(k-1)}$ at each step, producing an updated 
submix $\mathbf{s}^{(k)}$. This process repeats for $k = 1, \ldots, N$, 
with the final submix $\mathbf{s}^{(N)} = \hat{\mathbf{y}}_\text{mix}$.}
 \label{fig:framework_comp}
\end{figure}

Music production is a complex creative process requiring domain expertise, particularly in the mixing stage, where individual audio tracks are combined through a series of audio effects into a cohesive mixture balancing frequency content, spatial image, dynamics, and loudness~\cite{de2019intelligent}. To make mixing more accessible, the field of automatic music mixing (AMM) has emerged, with recent deep learning models demonstrating promising results~\cite{martinez2021deep, martinez2022automatic, steinmetz2021automatic, moliner2025automatic}.

Existing AMM approaches can be broadly categorized into two systems: implicit systems~\cite{martinez2021deep, martinez2022automatic}, which directly predict the output mixture end-to-end, and explicit systems~\cite{steinmetz2021automatic, moliner2025automatic}, which predict an intermediate representation of audio effects that is subsequently applied to the input tracks. While the former typically assumes a fixed number of input stems and fixed instrument types, the latter requires a predefined effects topology. That is, the type and ordering of audio effects processors must be specified at design time, regardless of whether they are implemented as differentiable DSP~\cite{engel2020ddsp} algorithms or pre-trained neural audio effects models. Despite these architectural differences, both types of systems rely on a parallelized architecture that processes all input tracks in a single pass.

While all existing approaches adopt a parallelized architecture that generates the final mixture in a single pass, an alternative paradigm remains unexplored: processing stems sequentially, one at a time. In this sequential paradigm, we name the new formulation as \textit{stem blending}: a conditional transformation that integrates a single stem into an existing submix\footnote{A \textit{stem} refers to an individual 
instrument track in a multitrack recording, such as vocals or drums. 
A \textit{submix} refers to a partial mixture formed by summing a 
subset of processed stems.}. In practice, this sequential paradigm is well-aligned with how professional mix engineers work, naturally exposing intermediate submixes at each step and allowing the mixing process to be inspected or initialized from any point. In this work, we propose a paradigm shift from single-pass to sequential mixing, and present stem blending as a principled reformulation of the AMM task, as shown in Figure~\ref{fig:framework_comp}.

Despite the intuitive motivation, realizing sequential stem blending in practice presents two key challenges. First, existing datasets are not directly applicable to stem blending: multitrack mixing datasets are designed for full mixture reconstruction rather than single-stem integration, while source separation datasets treat the mixture as a simple summation of stems without any mixing processing, providing no supervision signal for learning mixing transformations. Second, audio effects have conventionally been modeled in the audio 
domain, either through DSP algorithms~\cite{yu2025diffvox, 
lee2024searching, yeh2025ddsp} or waveform-domain neural networks 
such as convolutional~\cite{martinez2020deep, steinmetz2021efficient} 
or recurrent architectures~\cite{schmitz2018real, yeh2024hyper}, both requiring a predefined effects chain topology. This topology acts as a hard ceiling on the expressive power of the system: only the combinations of effects explicitly defined in the chain can be applied, 
and any processing outside this predefined set is simply not expressible, regardless of what the musical content demands.

To address these challenges, we present a latent flow matching model~\cite{lipman2022flow, liu2022flow} for sequential stem blending. For the data challenge, we propose a degradation-based data synthesis strategy that constructs paired training data by applying submix-informed or stem-informed audio degradations to simulate real-world mixing processing, rather than applying random audio effects. For the modeling challenge, we move away from explicit effects chain modeling and instead learn the stem blending transformation directly in the latent space using a flow matching model conditioned on the submix context. This allows the model to implicitly express arbitrary mixing transformations without being constrained to any predefined set of effects processors. At inference, stems are processed sequentially, with each step producing an intermediate submix that serves as the context for the next stem, enabling the model to handle an arbitrary number of input tracks.

We evaluate the proposed approach on a newly constructed stem blending benchmark as well as existing automatic music mixing benchmarks, through objective metrics and subjective listening tests. Experimental results demonstrate the effectiveness of the proposed stem blending model, and further reveal that the ordering of input stems influences the resulting mixing style, reflecting the context-dependent nature of the sequential blending process. Furthermore, our model naturally generalizes to the full AMM task by processing stems sequentially, suggesting that stem blending offers a viable and principled alternative formulation of automatic music mixing.

\section{Preliminaries on Flow Matching}\label{sec:background}


Flow matching defines a continuous mapping between a source distribution $p_0$ and a target distribution $p_1$ via an ordinary differential equation (ODE): $dz_t = v_{\theta}(z_t, t)dt$, where $v_{\theta}$ is a velocity field parameterized by a neural network with weights $\theta$ and $t \in [0, 1]$ is the continuous time step. We adopt rectified flow matching (RF)~\cite{liu2022flow}, which defines the forward process as $z_t = (1-t)z_0 + tz_1$, enforcing straight-line trajectories between samples $z_0 \sim p_0$ and $z_1 \sim p_1$ for more efficient integration. Under this formulation, the ground truth velocity is constant across time steps: $\hat{v}_t = z_1 - z_0$, and the network is trained by minimizing the expected MSE loss $\mathbb{E}_{t, z_0, z_1}\left[\| v_{\theta}(z_t, t) - \hat{v}_t \|^2\right]$. 


\section{Method}\label{sec:method}

\begin{figure*}[!th]
    \centerline{
        \includegraphics[width=\textwidth, height=8cm, keepaspectratio]{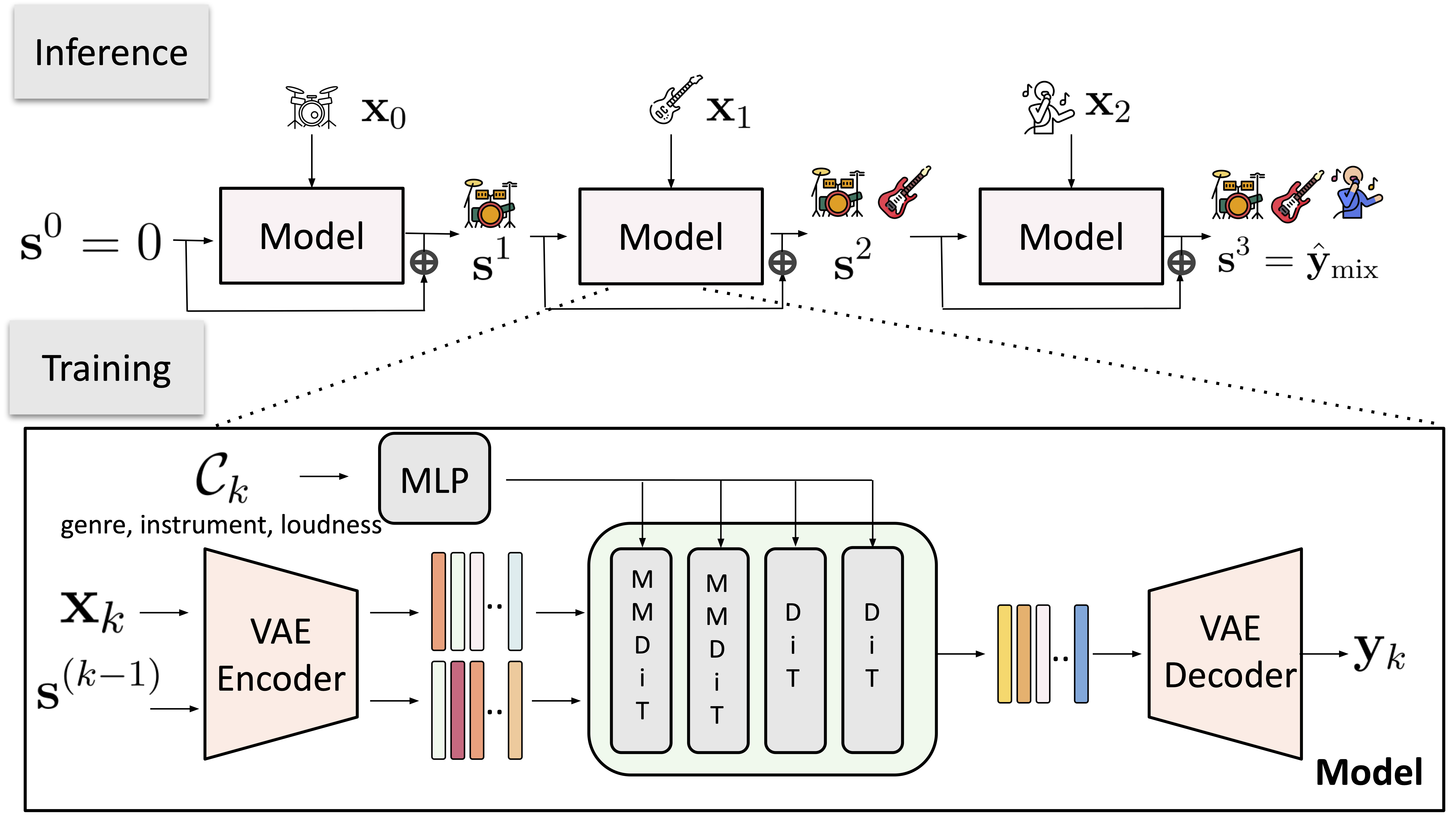}
    }
    \vspace{-2mm}
    \caption{Overview of the proposed system. \textit{Inference} (top): 
    stems are processed sequentially, where each stem $\mathbf{x}_k$ is 
    blended into the current submix $\mathbf{s}^{(k-1)}$ one at a time, 
    with the accumulated submix growing at each step until the final 
    mixture $\mathbf{s}^N = \hat{\mathbf{y}}_\text{mix}$ is obtained. 
    \textit{Training} (bottom): for each blending step, the unprocessed 
    stem $\mathbf{X}_k$ and the conditioning submix $\mathbf{s}^{(k-1)}$ 
    are encoded by a shared VAE encoder into latent sequences, where the 
    stem latent serves as the flow source $z_0$ and the submix latent is processed as a second modality via bidirectional 
    joint attention in the MMDiT blocks. The meta-conditioning set $\mathcal{C}_k$ 
    (genre, instrument, loudness) is processed by an MLP and injected into 
    all transformer blocks via AdaLN. The predicted latent 
    is decoded by the VAE decoder to produce the processed stem 
    $\mathbf{y}_k$.}
    \label{fig:model_arc}
\end{figure*}

\subsection{Problem Formulation}

Let $\mathbf{x}_i \in \mathbb{R}^{2 \times T}$ denote the $i$-th 
unprocessed stem represented as a stereo signal of $T$ samples. 
Existing AMM systems adopt a parallelized paradigm that maps a set of 
$N$ such stems $X = \{\mathbf{x}_i\}_{i=1}^N$ to a final mixture 
$\hat{\mathbf{y}}_{\text{mix}} \in \mathbb{R}^{2 \times T}$ in a 
single pass:
\begin{equation}
    \hat{\mathbf{y}}_{\text{mix}} = g\left(\{\mathbf{x}_i\}_{i=1}^N\right),
\end{equation}
where all stems are processed jointly in a single forward pass. In 
contrast, we propose a sequential paradigm that decomposes the mixing 
process into a series of stem blending steps. Given an initial submix 
$\mathbf{s}^{(0)} \in \mathbb{R}^{2 \times T}$ and an ordered set of 
stems $\{\mathbf{x}_k\}_{k=1}^N$, the mixture is constructed 
iteratively as:
\begin{equation}
    \mathbf{s}^{(k)} = \mathbf{s}^{(k-1)} + f\left(\mathbf{x}_k, 
    \mathbf{s}^{(k-1)}, \mathcal{C}_k\right), \quad k = 1, \ldots, N,
\end{equation}
where $\hat{\mathbf{y}}_{\text{mix}} = \mathbf{s}^{(N)}$. At each 
step, the conditioning submix $\mathbf{s}^{(k-1)}$ is held fixed and 
serves solely as context for blending $\mathbf{x}_k$, with only the 
predicted stem contributing to the updated submix $\mathbf{s}^{(k)}$. 
Here, $f$ denotes the stem blending transformation conditioned on the 
current submix $\mathbf{s}^{(k-1)}$ and a meta-conditioning set 
$\mathcal{C}_k$ comprising the integrated loudness of $\mathbf{x}_k$ 
and $\mathbf{s}^{(k-1)}$, as well as the genre and instrument type of 
$\mathbf{x}_k$.

\subsection{Overall Framework}

As shown in Figure~\ref{fig:model_arc}, our system consists of three main 
components. First, with a pre-trained encoder $\mathcal{E}$ and 
decoder $\mathcal{D}$, we encode the unprocessed stem $\mathbf{x}_k$ and 
the conditioning submix $\mathbf{s}^{(k-1)}$ into a latent 
space, yielding $z_0 = \mathcal{E}(\mathbf{x}_k)$ and 
$z_s = \mathcal{E}(\mathbf{s}^{(k-1)})$ respectively. Second, a 
rectified flow matching model learns the stem blending transformation 
$f$ in the latent domain, transporting $z_0$ toward the processed stem 
latent $z_1 = \mathcal{E}(\mathbf{y}_k)$ conditioned on $z_s$ and 
$\mathcal{C}_k$. The processed stem is then recovered as 
$\mathbf{y}_k = \mathcal{D}(\hat{z}_1)$. Third, a degradation-based 
data synthesis strategy constructs paired training data from existing 
multitrack mixing and source separation datasets. In the following 
subsections, we describe each component in detail.

\subsection{Stem Blending as Latent Transport}
We instantiate the stem blending transformation $f$ via rectified flow 
matching in the latent space of $\mathcal{E}$, learning a continuous 
transport from $z_0$ to $z_1$ conditioned on $z_s$ and $\mathcal{C}_k$. 
In the typical flow matching setting, $z_0$ is sampled from a Gaussian 
prior $\mathcal{N}(0, 1)$, $z_1$ represents the target data 
distribution, and the velocity field is conditioned on auxiliary 
information such as text prompts in text-to-music generation or query 
prompts in query-based source 
separation~\cite{prajwal2024musicflow, wu2026stemphonic, shi2025sam, 
yuan2025flowsep}. Since $\mathbf{x}_k$ and $\mathbf{y}_k$ share the 
same musical content but differ only in sonic characteristics, the flow 
is not required to generate new content from scratch, but rather to 
learn a transformation that modifies sonic characteristics while 
preserving the musical content. We therefore adopt a noise-free 
formulation inspired by~\cite{liu2025flowing, melechovsky2025sonicmaster}, 
initializing the flow directly from the stem latent $z_0 = 
\mathcal{E}(\mathbf{x}_k)$ rather than from Gaussian noise, where 
$z_1 = \mathcal{E}(\mathbf{y}_k)$ is the target processed stem latent. 
The velocity field $v_\theta$ is then trained by minimizing the flow 
matching objective:
\begin{equation}
    \mathcal{L}_{\text{FM}} = \mathbb{E}_{t, z_0, z_1}\left[\| 
    v_{\theta}(z_t, t, z_s, \mathcal{C}_k) - (z_1 - z_0) \|^2\right],
\end{equation}
where $z_t = (1-t)z_0 + tz_1$ is the linear interpolation between 
source and target latents at time step $t \in [0, 1]$, and $z_0$ 
refers to the degraded stem latent during training and the unprocessed 
stem latent at inference.

This design, however, requires careful treatment of the source 
distribution. As shown in previous work~\cite{liu2025flowing}, a 
deterministic encoder produces a point mass source distribution, which 
may cause the flow matching model to degenerate into deterministic 
regression. The source distribution $p(z_0)$ must therefore be 
regularized to preserve the necessary stochasticity. We address this 
naturally by adopting the Stable Audio Open VAE~\cite{evans2025stable} 
as our encoder, whose Gaussian-like latent distribution, enforced by 
the VAE training objective via KL divergence, provides the necessary 
stochasticity while preserving the acoustic structure of the input stem.

\subsection{Degradation-Based Data Synthesis}\label{dbds}

Since real recording sessions do not capture the submix state at each intermediate blending step, obtaining ground truth for these stages is infeasible in practice. We therefore construct training data by simulating only the final blending step, where the submix $\mathbf{s}^{(N-1)}$ aggregates all but one of the $N$ stems in a song.

For MedleyDB~\cite{bittner2014medleydb}, where raw and wet stem pairs are directly available, we iterate over each stem $k$ of a song to form a triplet: its raw version as $\mathbf{x}_k$, its wet version as $\mathbf{y}_k$, and the sum of all other wet stems as the submix $\mathbf{s}^{(N-1)}$. For MoisesDB~\cite{pereira2023moisesdb}, only wet stems are available, so we simulate $\mathbf{x}_k$ by applying degradations to $\mathbf{y}_k$. Each degradation mode is designed as the inverse of a common mixing engineer decision, simulating the specific sonic problems a mix engineer would identify and correct, rather than applying random effects. Spectral degradations simulate bad mixing scenarios via a parametric equalizer with five modes: \textit{masking boost} boosts frequencies 
occupied by the submix; \textit{over-cut} cuts the stem's own prominent bands; \textit{low-end mud} applies a heavy low-shelf boost and high cut; \textit{harshness} adds a narrow boost in the 2--5~kHz region; and \textit{blend} combines submix-informed boost with stem-informed cut. The first two modes are conditioned on the submix and stem spectral profiles respectively, reflecting real-world mixing decisions. 
Additionally, room reverberation is randomly applied to simulate stems recorded in reverberant or acoustically untreated spaces, using simulated room impulse responses of varying room sizes and absorption characteristics~\cite{scheibler2018pyroomacoustics}.

\subsection{Sequential Inference}

A key advantage of the sequential paradigm is that the model naturally 
supports interactive workflows. Since each stem is processed 
conditioned on the current submix, users can provide their own submix 
at any point, blend a specific stem into an existing mix, or inspect 
intermediate outputs at each step. This flexibility is
unavailable in single-pass approaches.

For automatic music mixing, the sequential process must be 
initialized with an initial submix $\mathbf{s}^{(0)}$. We initialize $\mathbf{s}^{(0)}$ with a zero-valued latent vector and process the first stem to obtain an initial result, after which remaining stems are processed sequentially. We investigate two ordering strategies for the remaining stems. The first processes stems in a \textit{random} order, serving as an ordering-agnostic baseline. The second follows a \textit{domain-knowledge} ordering based on the tonal and rhythmic dependency hierarchy of instruments: \texttt{drums} $\rightarrow$ 
\texttt{bass} $\rightarrow$ \texttt{guitar} $\rightarrow$ 
\texttt{keys} $\rightarrow$ \texttt{strings} $\rightarrow$ 
\texttt{vocals} $\rightarrow$ \texttt{other}. This ordering reflects one common practice of establishing rhythmically and tonally foundational stems first, allowing melodically dependent stems to adapt to the established mix context.

\subsection{Implementation Details}

\noindent \textbf{Model Architecture \& Training.} The flow matching model $v_\theta$ employs a hybrid architecture combining Multimodal Diffusion Transformer (MMDiT)~\cite{esser2024scaling} blocks with subsequent Diffusion Transformer (DiT)~\cite{peebles2023scalable} layers, specifically 2 MMDiT dual-stream blocks and 2 DiT 
single-stream layers, operating on the latent space of the Stable Audio Open VAE~\cite{evans2025stable}. The submix latent $z_s$ is projected to the joint attention dimension and processed as a second modality via bidirectional joint attention in the MMDiT blocks, while the meta-conditioning set $\mathcal{C}_k$ is encoded via learned embedders and fused into a global conditioning vector injected via AdaLN. To enable classifier-free guidance~\cite{ho2022classifier}, the submix audio latent and submix loudness are independently masked during training with a drop probability of 0.1, while genre, 
instrument type, and raw stem loudness are never masked. The model is trained for 30 epochs using AdamW with $\beta_1 = 0.9$, $\beta_2 = 0.95$, a learning rate of $3 \times 10^{-4}$ with cosine decay and 500 warmup steps, a batch size of 128, and 10-second audio segments 
at 44.1 kHz on a single NVIDIA RTX 4090 GPU, with 10 inference steps.

\noindent \textbf{Dataset.} For training, we use MedleyDB v1~\cite{bittner2014medleydb}, comprising 122 songs with raw and wet stem pairs, and MoisesDB~\cite{pereira2023moisesdb}, comprising 216 songs after reserving 24 for the stem blending benchmark. Since both 
datasets use heterogeneous instrument naming conventions, we map all stem labels to seven canonical instrument groups, namely \texttt{vocals}, \texttt{drums}, \texttt{bass}, \texttt{guitar}, \texttt{keys}, \texttt{strings}, and \texttt{other}, used as the 
instrument type conditioning in $\mathcal{C}_k$. Audio segments of 10 seconds are extracted with a 50\% hop overlap, discarding segments with low RMS or active frame ratio below 0.4. For the stem blending benchmark, we randomly sample 300 segments from the 24 held-out 
MoisesDB songs. For automatic music mixing evaluation, we use MedleyDB v2, randomly sampling 364 segments; note that MedleyDB v1 and v2 share no overlapping songs, ensuring a clean separation between training and evaluation data.
\section{Experimental Setup}\label{sec:experiments}

\begin{table*}[!ht]
  \centering
  \renewcommand{\arraystretch}{1}
  \setlength{\tabcolsep}{5pt}
  \begin{tabular}{llcrrccc}
    \toprule
    & \multirow{2}{*}{Model} & \multirow{2}{*}{Type} & \multicolumn{2}{c}{KAD $\downarrow$} & \multicolumn{3}{c}{FD $\downarrow$} \\
    \cmidrule(lr){4-5} \cmidrule(lr){6-8}
    & & & FxEnc++ & CLAP & RMS & TB & CF \\
    \midrule
    \multicolumn{8}{l}{\textit{Stem Blending} {\small (1-stem + submix, MoisesDB)}} \\
    \midrule
    & Raw-mix          & --   & 1.42  & 1.61  & 4.86e+02 & 2.59e+03 & 3.39 \\
    \cmidrule(lr){2-8}
    & DMC~\cite{steinmetz2021automatic}              & Deterministic & 46.45 & 1.76  & 1.71e+03 & 1.03e+03 & 1.25 \\
    & DMC$^\dagger$~\cite{steinmetz2021automatic}     & Deterministic & 24.39 & 7.59  & 6.57e+02 & 9.24e+02 & 4.73 \\
    & MEGAMI~\cite{moliner2025automatic}          & Generative & 11.67 & 8.18  & 2.60e+03 & 6.42e+03 & 2.94 \\
    & MEGAMI$^\dagger$~\cite{moliner2025automatic}  & Generative & 7.06  & 7.72  & 2.20e+03 & 1.07e+04 & 6.45 \\
    \cmidrule(lr){2-8}
    & Proposed         & Generative & \textbf{--0.01} & \textbf{--0.07} & \textbf{5.37e+00} & \textbf{3.25e+01} & \textbf{0.03} \\
    \midrule
    \multicolumn{8}{l}{\textit{Automatic Music Mixing} {\small ($K$-stem, MedleyDB v2)}} \\
    \midrule
    & Raw-mix          & --   & 37.02 & 9.82  & 9.26e+02 & 7.69e+01 & \textbf{0.19} \\
    \cmidrule(lr){2-8}
    & DMC~\cite{steinmetz2021automatic}               & Deterministic & 39.08 & 11.61 & 6.24e+01 & \textbf{4.21e+01} & 0.27 \\
    & MEGAMI~\cite{moliner2025automatic}            & Generative & \textbf{7.99}  & 10.11 & 1.28e+02 & 1.06e+02 & 0.84 \\
    \cmidrule(lr){2-8}
    & Proposed-Random  & Generative & 32.95 & 6.66  & 4.21e+01 & 2.84e+02 & 1.01 \\
    & Proposed-Domain  & Generative & 31.01 & \textbf{6.52}  & \textbf{3.71e+01} & 3.24e+02 & 0.81 \\
    \bottomrule
  \end{tabular}
  \caption{Objective evaluation results on stem blending and automatic 
    music mixing benchmarks. The proposed model is evaluated under two 
    ordering strategies: random (\textit{Proposed-Random}) and 
    domain-knowledge (\textit{Proposed-Domain}). For DMC and MEGAMI, 
    $^\dagger$ denotes the re-blending variant where the predicted stem 
    is summed with the original submix to enable stem blending evaluation. 
    TB: tonal balance, CF: crest factor. \textbf{Bold} indicates the best 
    result in each scenario.}
  \label{tab:main_results}
\end{table*}

\subsection{Evaluation Scenarios}
We evaluate on two benchmarks. For the \textit{stem blending benchmark}, a single stem is blended into the submix $\mathbf{s}^{(N-1)}$ formed by the remaining stems, and the resulting mixture is evaluated against the professionally mixed reference. Since the benchmark is constructed using the same degradation-based strategy described in Section~\ref{dbds}, it provides a controlled setting to directly measure whether the model learns the intended blending transformation, with generalization to real-world scenarios assessed via perceptual evaluation on out-of-distribution stems (Section~\ref{sec:se}). For the \textit{automatic music mixing benchmark}, all stems are processed sequentially to produce the final mixture, evaluating the model's generalization to the full AMM task.


\subsection{Baselines}
We compare against three baselines. \textit{Raw-mix} sums all unprocessed stems without any mixing processing. \textit{DMC}~\cite{steinmetz2021automatic} predicts parameters for a 
fixed effects chain comprising gain, equalization, compression, panning, and reverb from the \texttt{dasp-pytorch} package~\cite{dasp}; we use the advanced version from 
Diff-MST~\cite{vanka2024diff} without the mixing style transfer mechanism, retrained on MedleyDB v1. \textit{MEGAMI}~\cite{moliner2025automatic} is a generative framework 
operating in an effect embedding space, evaluated using its publicly available checkpoint\footnote{\url{https://github.com/SonyResearch/MEGAMI}}. Both baselines produce individual stem outputs, enabling evaluation under both scenarios. For DMC and MEGAMI, we additionally evaluate a re-blending variant (denoted $\dagger$) where 
only the predicted stem is summed with the original submix, preventing parallelized models from inadvertently modifying the submix context and enabling further comparison under the stem blending scenario.

\subsection{Evaluation Metrics}
Since mixing is a one-to-many task, pairwise metrics comparing outputs to a single reference are not well-suited for evaluation~\cite{moliner2025automatic}. We therefore measure the 
distributional distance between system outputs and professionally mixed reference recordings using two metric categories.

\noindent \textbf{Embedding-based metrics.} We use Kernel Audio Distance (KAD)~\cite{chung2025kad} based on Maximum Mean Discrepancy, which overcomes the limitations of FAD~\cite{kilgour2018fr} while demonstrating stronger alignment with human perceptual judgments. KAD is computed with two embedding spaces: CLAP~\cite{wu2023large} 
embeddings capture overall perceptual quality, while FxEncoder++~\cite{yeh2025fx} captures mixing style similarity independent of musical content, together providing complementary 
perspectives on the generated mixture.

\noindent \textbf{Signal-level metrics.} We compute the Fréchet Distance (FD) between distributions of three signal-level features: tonal balance (frequency content distribution), RMS (time-varying energy), and crest factor (peak-to-RMS ratio), providing fine-grained analysis of mixing quality that complements the embedding-based metrics.

\subsection{Perceptual Evaluation Setup}\label{sec:se}
To evaluate generalization beyond the training distribution, we design a perceptual evaluation using stems generated by the stem generation model from Moises\footnote{\url{https://moises.ai/}}. We select three songs spanning rock, pop, and jazz. For each song, we apply source separation to obtain five stems (\texttt{vocals}, \texttt{drums}, \texttt{bass}, \texttt{guitar}, \texttt{other}), then independently regenerate \texttt{drums}, \texttt{bass}, and \texttt{guitar}, each conditioned on the original versions of the remaining four stems (e.g., \texttt{drums} is regenerated given the original \texttt{vocals}, \texttt{bass}, \texttt{guitar}, and \texttt{other}). This avoids overlap with the training distribution while reflecting a realistic production workflow where stems are generated and mixed.

We construct two evaluation scenarios. In \textit{stem blending} (STB), one of the three regenerated stems is randomly selected and blended into the submix of original remaining stems, isolating blending behavior on out-of-distribution input. In \textit{automatic music mixing} (AMM), all three generated stems are mixed with the original \texttt{vocals} and \texttt{other} into a coherent full mixture.

We conduct a MUSHRA-style listening test~\cite{schoeffler2018webmushra} on 6 samples (one per song per scenario), in which 18 participants with musical background rate perceptual quality from 0 (Bad) to 100 (Excellent). Samples from both scenarios are presented together without scenario labels to avoid listener bias, and four systems are compared: Raw-mix, DMC, MEGAMI, and the proposed model. We additionally evaluate all samples using the Meta Audiobox Aesthetics model~\cite{tjandra2025meta} to obtain model-based production quality (PQ) scores as a complementary automatic measure\footnote{Meta Audiobox Aesthetics was trained on mono, 16 kHz audio, so we report its scores only as a bandwidth-limited proxy for music-mixing quality.}. Results are reported in Section~\ref{sec:percep_eval}.

\section{Objective Results}\label{sec:obj_results}

While music mixing ultimately involves subjective judgment, the objective metrics in this section offer useful insights into the behavioral properties of each system, complementing the perceptual study in Section~\ref{sec:percep_eval}.

\noindent \textbf{Stem Blending.}
Table~\ref{tab:main_results} presents the stem blending results. Notably, \textit{Raw-mix outperforms all AMM baselines despite applying no processing}, revealing that parallelized models have no mechanism to distinguish whether an input is already well-processed, inadvertently modifying the submix context and disrupting its musical coherence. Even with the re-blending adaptation ($\dagger$), both DMC and MEGAMI still perform poorly, confirming that parallelized approaches are inherently not designed for the stem blending formulation. In contrast, the proposed model achieves substantially lower distributional distances across all metrics, with near-zero KAD values indicating that the output distribution closely matches the reference~\cite{chung2025kad}.

\noindent \textbf{Automatic Music Mixing.}
Table~\ref{tab:main_results} presents the AMM results. The proposed model outperforms DMC and Raw-mix on CLAP KAD and RMS FD, providing empirical evidence that sequential stem blending generalizes to the full AMM task \textit{despite being trained solely on single-stem blending}. Proposed-Domain consistently outperforms Proposed-Random, suggesting that domain-knowledge ordering yields more coherent final mixtures. While MEGAMI achieves better FxEnc++ KAD, likely due to its explicit FxEncoder++ embedding generation during inference, our model achieves the best CLAP KAD, reflecting a tradeoff between mixing style similarity and overall perceptual quality. Beyond comparative ranking, these metrics also expose characteristic behaviors of our model: Raw-mix achieves the lowest crest factor FD, suggesting that unprocessed stems already exhibit a dynamic range distribution close to professional mixes, while both proposed variants show higher crest factor FD, indicating that the latent blending transformation introduces dynamic range changes that deviate from the reference distribution. Similarly, both implicit transformation methods show higher tonal balance FD compared to DMC, suggesting that our model tends to reshape spectral content differently from the reference distribution.

\begin{table}[!t]
  \centering
  \renewcommand{\arraystretch}{1}
  \setlength{\tabcolsep}{5pt}
  \begin{tabular}{lcccc}
    \toprule
    \multirow{2}{*}{Model} & \multicolumn{3}{c}{PQ $\uparrow$} \\
    \cmidrule(lr){2-4}
    & STB & AMM & Overall \\
    \midrule
    Raw-mix  & 7.741 & 7.733 & 7.737 \\
    \midrule
    DMC~\cite{steinmetz2021automatic}       & 7.800 & 7.667 & 7.734 \\
    MEGAMI~\cite{moliner2025automatic}    & 7.506 & 7.799 & 7.653 \\
    Proposed & \textbf{8.104} & \textbf{7.837} & \textbf{7.971} \\
    \bottomrule
  \end{tabular}
  \caption{Production quality (PQ) scores evaluated on stem blending (STB) and automatic music mixing (AMM) benchmarks.}
  \label{tab:pq}
\end{table}

\begin{figure}[!t]
 \centerline{
 \includegraphics[width=1.0\columnwidth, keepaspectratio]{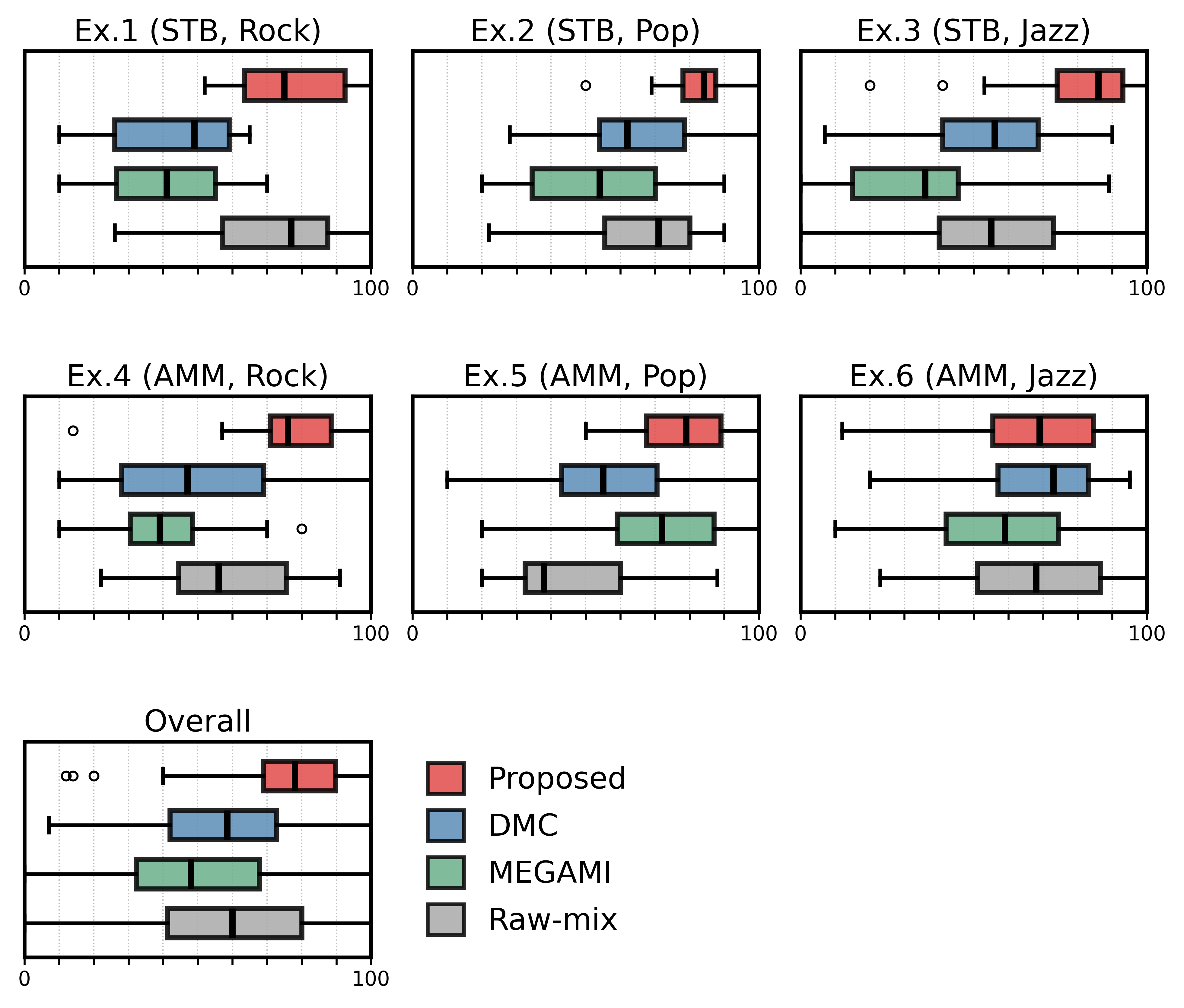}
 }
 \vspace{-2mm}
 \caption{Subjective results from $18$ participants across Automatic Music Mixing (AMM) and Stem Blending (STB) scenarios.}
 \label{fig:sub}
\end{figure}

\section{Perceptual Evaluation}\label{sec:percep_eval}

\noindent \textbf{Meta Audiobox Aesthetics.}
Table~\ref{tab:pq} presents the production quality (PQ) scores from the Meta Audiobox Aesthetics model~\cite{tjandra2025meta} on both scenarios. The proposed model achieves the highest score in both and the highest overall mean. On STB, it shows the largest margin over all baselines, while DMC ranks second and MEGAMI scores below Raw-mix, suggesting that generative effect embedding approaches are less suited for single-stem blending. On AMM, DMC falls below Raw-mix, which we attribute to its sensitivity to out-of-distribution stems, while MEGAMI recovers to rank second, aligning with its design as a full-mix system.

\noindent \textbf{Subjective Listening Tests.}
Figure~\ref{fig:sub} presents MUSHRA score distributions from 18 participants across six examples covering STB (Ex.~1--3) and AMM (Ex.~4--6) scenarios across Rock, Pop, and Jazz. Surprisingly, despite the deviations in crest factor and tonal balance seen in the objective metrics, the proposed model achieves the highest median score in five of six examples and overall, suggesting that perceptual evaluation involves additional factors such as listener taste and musical context that distributional metrics may not fully capture.

Examining individual systems, Raw-mix shows high variance, reflecting the inconsistent nature of unprocessed stems whose perceived quality depends heavily on how well the source recordings happen to balance with one another. AMM baselines can in turn hurt quality when processing is inappropriate, as evidenced by cases where DMC and MEGAMI score below Raw-mix, echoing our earlier observation that parallelized models lack a mechanism to recognize when a stem is already well-suited to its mix context. In contrast, the proposed model achieves consistently higher scores with a tighter distribution, which we attribute to its stem blending formulation: by integrating one stem at a time into a fixed submix, the model treats the existing context as an acoustic anchor and adapts the incoming stem accordingly, gently avoiding the over-processing observed in the parallelized baselines. Since these evaluations use stems generated by an out-of-distribution stem generation model, the consistency of these results across both scenarios further suggests that the degradation-based training strategy captures generalizable mixing behavior beyond the specific degradation modes seen during training.

\section{Conclusion}\label{sec:conclusion}
We proposed sequential stem blending as a principled reformulation of automatic music mixing, demonstrating that existing parallelized approaches are inherently not designed for this task while our model achieves strong stem blending performance and competitive results on the full AMM task. Beyond performance, the sequential paradigm naturally supports interactive workflows where users blend specific stems into an existing mix or inspect intermediate outputs, and reframes stem ordering, often treated as a nuisance variable~\cite{steinmetz2021automatic, moliner2025automatic}, as a lightweight, training-free mechanism for mixing style control. Full AMM performance does not yet match state-of-the-art systems on tonal balance and mixing style similarity, which we partly attribute to the degradation strategy simulating only the final blending step, leaving the model unexposed to early steps with sparse submixes. We view stem blending as a complementary formulation of AMM and hope it inspires future work on sequential and interactive mixing systems.

\section{Acknowledgements}
The work is supported by grants from Google Asia Pacific, the National Science and Technology Council of Taiwan (NSTC 114-2628-E-002-013-MY3), and the Ministry of Education (MOE) of Taiwan (for Taiwan Centers of Excellence in Artificial Intelligence). 

\bibliography{ISMIRtemplate}

%
%
%
%

\end{document}